\documentclass[aps,prl,twocolumn,groupedaddress,amsmath,amssymb,amsfonts,graphics]{revtex4}

\usepackage{graphicx}
\usepackage{bm}
\usepackage{bbold}
\usepackage{dcolumn}
\usepackage{booktabs}

\begin{document}

\title{Comment on ``High Precision Measurement of the Thermal Exponent
for the three-dimensional XY Universality Class'' }

\author{K.\ S.\ D.\ Beach}
\affiliation{Department of Physics, Boston University, 590 Commonwealth Avenue, Boston, MA 02215}

\date{July 18, 2005}

\begin{abstract}
A recent paper [Burovski~\emph{et al}., \texttt{cond-mat/0507352}] reports
on a new, high-accuracy simulation of the classical $\phi^4$ model 
(in the three-dimensional XY universality class). 
The authors claim that a careful scaling analysis of their data gives 
$\nu = 0.6711(1)$ for the thermal critical exponent. If correct, this would neatly resolve
the discrepancy between numerical simulations and experiments on ${}^4$He.
There is reason, however, to doubt the accuracy of the result.
A re-analysis of the data yields a significantly higher value of $\nu$,
one that is consistent with other Monte Carlo studies.
\end{abstract}

\maketitle

Universality is an elegant and powerful concept, but as the basis for any method of
data analysis it offers many dangers to the practitioner. For example, it is notoriously 
difficult to extract critical exponents from data on finite systems.
Such an analysis is exquisitely sensitive to the choice of scaling form,
to the number of fitting parameters, and to the range of system sizes included
in the fit. Worse, it is quite difficult to quantify
the uncertainties associated with these factors, so there is a tendency 
to overstate the accuracy of measured exponents. Reference~\onlinecite{Burovski05}
may suffer from this very problem.

After repeated re-analysis of their experiments on the
superfluid transition of ${}^4$He in microgravity~\cite{Lipa96,Lipa00,Lipa03},
Lipa~\emph{et al}.\ have concluded that the best experimental value
of the thermal critical exponent is $\nu_{\text{exp}} = 0.6709(1)$. This
value is not in good agreement with numerical simulations of other
models also believed to be in the three-dimensional XY universality class.
Two recent Monte Carlo studies~\cite{Hasenbusch99,Campostrini01} 
found $\nu_{\text{mc}} = 0.6723(3)(8)$ and $0.67155(27)$.

In an effort to settle the controversy, Burovski~\emph{et al}.\ studied the 
classical $\phi^4$ model~\cite{Burovski05}
\begin{equation} \label{EQ:phi4model}
\frac{H}{T} = -t\sum_{\langle ij \rangle} \phi_{i}^{*} \phi_{j} +
\frac{U}{2} \sum_{i} |\phi_{i}|^4 - \mu \sum_{i} | \phi_{i} |^2
\end{equation}
by expanding the partition function in series and sampling
the various terms stochastically. They considered two
critical points---
\begin{alignat*}{3}
\text{point A}&: \quad & t &= -0.0795548(1),    & \quad U &= 0.4101562(14) \\
\text{point B}&:            & t & = -0.07142883(7), &            U &= 0.3605750(8)
\end{alignat*}
---the second chosen so as to minimize the subleading corrections to finite-size
scaling (although one wonders if this helps or simply makes it harder to pick out the
subleading corrections when it comes time to fit the data).

Burovski \emph{et al}.\ computed the superfluid stiffness $\rho_{\text{s}}$ 
and two of its $t$ derivatives for a range of linear systems 
sizes $L = 4, \ldots, 96$. Their results for $R' = \partial(\rho_s L)/\partial t$, 
evaluated at the two critical points, are reproduced in Table~\ref{TAB:Rprime}.
Each of the entries corresponds to $5\times 10^8$ Monte Carlo sweeps. 

\begin{table}[!tb]
\caption{ \label{TAB:Rprime} Derivative of the superfluid stiffness scaling function
at critical points A and B computed via Monte Carlo.}
\begin{ruledtabular}
\begin{tabular}{@{\qquad}d|d@{\qquad}|d@{\qquad}|@{\qquad}}
\multicolumn{1}{c|}{}
& \multicolumn{2}{c}{$R' = \partial(\rho_{\text{s}}L)/ \partial t \rvert_{t=t_{\text{c}}}$} \\
L & \multicolumn{1}{c}{data set A} & \multicolumn{1}{l}{\qquad\! data set B} \tabularnewline \hline
4 & 2.0329(9)  & 1.9907(3) \tabularnewline
5 & 2.8414(5)  & 2.7843(7) \tabularnewline
6 & 3.7316(4)  & 3.6586(1)\footnotemark[1] \tabularnewline
7 & 4.6955(5)  & 4.6064(5) \tabularnewline
8 & 5.7289(4)  & 5.6221(9) \tabularnewline
9 & 6.8265(7)  &  \tabularnewline
10 & 7.9848(9) & 7.840(1) \tabularnewline
11 & 9.2031(13) &  \tabularnewline 
12 & 10.474(2) & 10.286(1) \tabularnewline
16 & 16.074(4) & 15.789(2) \tabularnewline
20 & 22.403(4) & 22.020(3) \tabularnewline
24 & 29.396(7) & 28.897(3) \tabularnewline
32 & 45.095(13) & 44.36(1) \tabularnewline
48 & 82.48(3)& 81.15(2) \tabularnewline
64 &  & 124.57(7)\footnotemark[1]  \tabularnewline
96 & 231.56(17) & 
\footnotetext[1]{points of disagreement between Fig.~\ref{FIG:data-fitting}
of this paper and Fig.~3 of Ref.~\onlinecite{Burovski05}.}
\end{tabular}
\end{ruledtabular}
\end{table}

They simultaneously fit the two data sets, assuming a scaling form
\begin{equation} \label{EQ:Rprime}
R' = (\text{const.}) L^{1/\nu}\bigl[1+ (\text{const.}) L^{-\omega}\bigr].
\end{equation}
The constants were allowed to take on different values at 
critical points A and B, whereas the exponents $\nu$ and $\omega$ were 
assumed to be universal ($\nu = \nu_{\text{A}} = \nu_{\text{B}}$, \emph{etc}.).
On the basis of a known irrelevant exponent,
the fit was constrained such that $\lvert \omega - 0.795\rvert < 0.03$;
data for system sizes below $L_{\text{cutoff}} = 12$ were discarded.
Final fit values of $\omega = 0.796(3)$ and $\nu = 0.6711(1)$ were reported.

\begin{figure}[!t]
\includegraphics[scale=0.95]{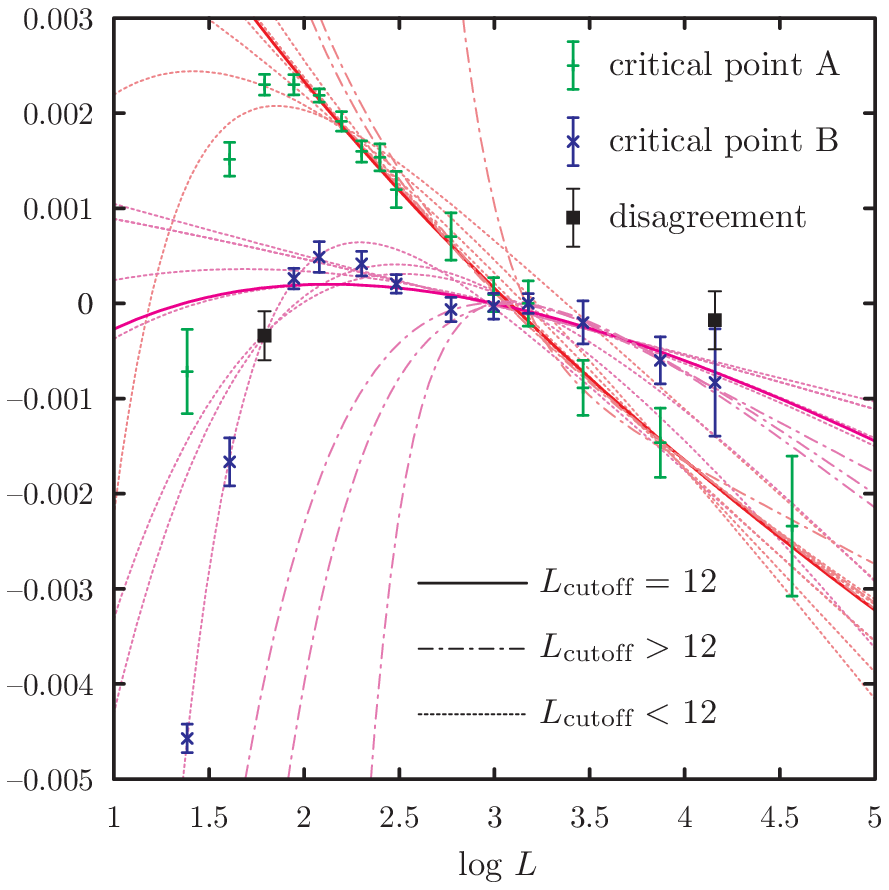}
\includegraphics[scale=0.95]{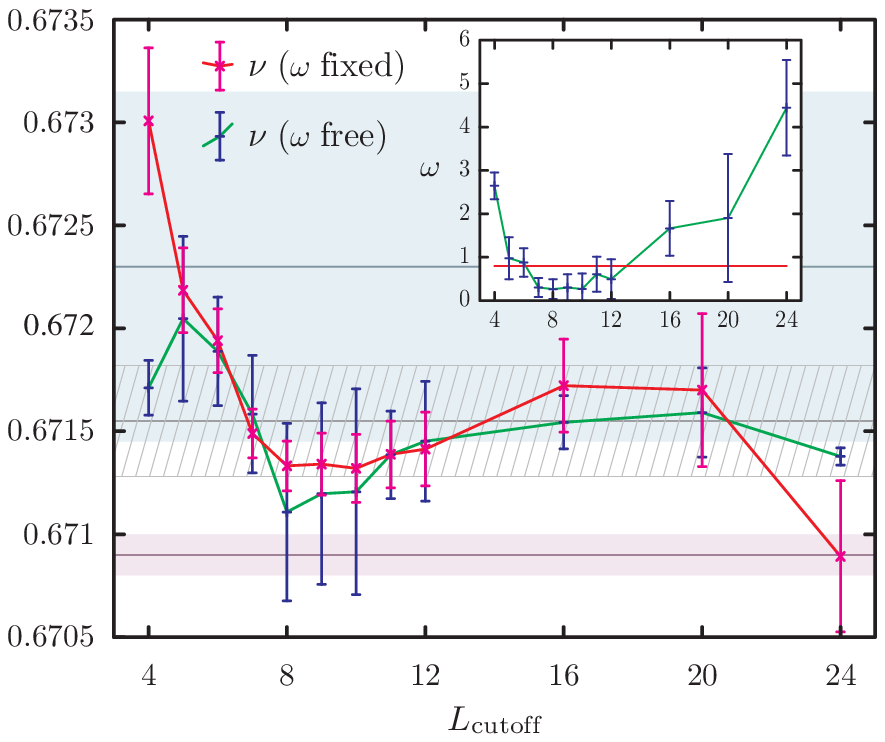}
\caption{ \label{FIG:data-fitting} (Top panel)
The data from Table~\ref{TAB:Rprime} are plotted exactly as in
Fig.~3 of Ref.~\onlinecite{Burovski05}. 
The two black squares mark the data points in
Fig.~3 (at $L=6$ and $L=64$) that do not coincide with those shown here.
The solid lines denote the best fits 
for critical points A (red) and B (pink) with $L_{\text{cutoff}} = 12$
and $\omega = 0.796$ fixed. Best fits for $L_{\text{cutoff}} = 4,5,6,7,8,9,10,11$
are drawn dotted and those for $L_{\text{cutoff}} = 16,20,24$ are drawn dot-dashed.
(Bottom panel) The optimal $\nu$---computed with both $\omega$ held fixed
and $\omega$ left free to vary---is plotted as a function of the lower size cutoff.
The corresponding values of $\omega$ are shown in the figure inset.
The violet band indicates one standard deviation above and below
$\nu_{\text{exp}}$ from Ref.~\onlinecite{Lipa03}.
The Monte Carlo results from Refs.~\onlinecite{Hasenbusch99} 
and Ref.~\onlinecite{Campostrini01} are marked by the cyan and hatched regions.
}
\end{figure}

Is this value of $\nu$ convincing? Such a high degree of confidence in 
the fourth digit is probably not warranted.
It is clear even to the eye---in Figs.~2 and 3 of Ref.~\onlinecite{Burovski05}---that 
the data allows considerable leeway in the slope and offset of the fit.
Moreover, two aspects of the data analysis are troubling:
\begin{enumerate}
\item It is not particularly useful to
restrict the value of the scaling exponent. While it is true that the
lowest-order subleading correction arises from an irrelevant scaling field
with exponent $\omega_{\text{irr}} \sim 0.8$, that correction coexists with other
contributions analytic in $L^{-1}$
[which are not explicitly included in Eq.~\eqref{EQ:Rprime}].
In practice, $\omega$, as it appears in Eqs.~(4) and (6) of Ref.~\onlinecite{Burovski05},
is an \emph{effective} exponent that approximates the subleading
behaviour over some range of $L$.~\cite{Beach05}
\item More important, the authors have failed to quantify the effect of $L_{\text{cutoff}}$
on their fit. If, as I believe is true in this case, the optimal $\nu$ depends
sensitively on the choice of the lower size cutoff, then some convincing
criterion must be advanced to justify choosing one value of $L_{\text{cutoff}}$ over another.
\end{enumerate}

In the top panel of Fig.~\ref{FIG:data-fitting}, I have replotted the superfluid
stiffness data of Burovski~\emph{et al}.\ alongside my own best fits. These
fits were generated using exactly the procedure outlined in Ref.~\onlinecite{Burovski05}
for a range of $L_{\text{cutoff}}$, with and without constraints on $\omega$.
The large variability in $\nu$, as seen in the bottom panel of Fig.~\ref{FIG:data-fitting},
suggests that an uncertainty of $\pm 0.0001$ is overly optimistic.
Moreover, contrary to the authors primary claim, it appears that
(for $6 \lesssim L_{\text{cutoff}} \lesssim 20$) $\nu$
is largely consistent with the result of Campostrini \emph{et al}.\ and not
significantly closer to the experimental value. 
There is little compelling evidence for a value as low as 0.6711.

It is important to note that there is a curious disagreement 
(in two data points from the B data set ) between Fig.~\ref{FIG:data-fitting} 
of this paper and Fig.~3 of Ref.~\onlinecite{Burovski05}.
In Fig.~3, the $L=6$ data point has the same position, but a larger errorbar; the
$L=64$ data point is placed considerably higher. The origin of this discrepancy
is somewhat mysterious. It may simply be a consequence of minor typographical
errors in Table~\ref{TAB:Rprime}. In that case, Burovski \emph{et al}.\ are fitting
to slightly different data set than I am. Nonetheless, it is telling that minor changes on the 
order of one or two standard deviations to two of twenty-eight data points could have 
such a disruptive effect. It suggests that $\nu = 0.6711(1)$ is an unrealistic
estimate of the thermal critical exponent.

\end{document}